%% file: main.tex
\documentclass[preprint,12pt]{elsarticle}

\usepackage{float}
\usepackage{amsfonts}
\usepackage{amsmath}
\usepackage{amssymb}
\usepackage{hyperref}
\usepackage{url}
\usepackage{graphicx}
\usepackage{xcolor}
\usepackage{tikz}
\usetikzlibrary{automata,arrows,positioning,calc}
\usepackage{pgfplots}
\pgfplotsset{compat=1.15}
\usepackage{algorithm}
\usepackage{algpseudocode}
\usepackage{balance}
\usepackage{makecell}
\journal{Internet of Things}

\begin{document}

\begin{frontmatter}



\title{Availability Evaluation of IoT Systems with Byzantine Fault-Tolerance for Mission-critical Applications}


\author[inst1]{Marco~Marcozzi}
\author[inst2]{Orhan~Gemikonakli}
\author[inst3]{Eser~Gemikonakli}
\author[inst4]{Enver~Ever}
\author[inst1]{Leonardo~Mostarda}

\affiliation[inst1]{organization={Computer Science Division, University of Camerino},
            addressline={Via~Madonna~delle~Carceri,~7}, 
            city={Camerino~(MC)},
            postcode={62032}, 
            country={Italy}}

\affiliation[inst2]{organization={Faculty of Engineering, Final International University},
            city={Kyrenia},
            country={Cyprus}}

\affiliation[inst3]{organization={Department of Computer Engineering, Faculty of Engineering, University of Kyrenia},
            addressline={Mersin 10}, 
            city={Girne},
            country={Turkey}}

\affiliation[inst4]{organization={Computer Engineering, Middle East Technical University Northern Cyprus Campus},
            addressline={Mersin 10}, 
            city={Guzelyurt},
            postcode={99738}, 
            country={Turkey}}

\begin{abstract}
Byzantine fault-tolerant (BFT) systems are able to maintain the availability and integrity of IoT systems, in the presence of failure of individual components, random data corruption or malicious attacks. Fault-tolerant systems in general are essential in assuring continuity of service for mission-critical applications. However, their implementation may be challenging and expensive. In this study, IoT Systems with Byzantine Fault-Tolerance are considered. Analytical models and solutions are presented as well as a detailed analysis for the evaluation of the availability. Byzantine Fault Tolerance is particularly important for blockchain mechanisms, and in turn for IoT, since it can provide a secure, reliable and decentralized infrastructure for IoT devices to communicate and transact with each other.
A continuous-time Markov chain is used to model the IoT systems with Byzantine Fault-Tolerance where the breakdown and repair times follow exponential distributions, and the number of the Byzantine nodes in the network follows various distributions. The presented numerical findings demonstrate the relationship between the number of servers in the system, the proportion of honest users, and the overall availability. Based on the model, it can be inferred that the correlation between the scale of the system (servers) and network availability is non-linear.
\end{abstract}




\end{frontmatter}


\section{Introduction}\label{sec:introduction}
IoT (Internet of Things) devices are often deployed in distributed and decentralized environments, where a large number of devices need to communicate and coordinate with each other to perform various tasks. BFT (Byzantine Fault Tolerance) systems can benefit IoT, particularly in terms of improved reliability and fault tolerance. BFT systems are designed to ensure the system's reliability and fault tolerance in a distributed environment. They can tolerate a certain number of faulty or malicious nodes in the network without compromising the system's overall performance. This feature can be particularly beneficial in IoT systems where a large number of devices need to communicate and coordinate with each other, and the failure of a single node can have a significant impact on the entire system's performance \cite{fernandez2018review}.

Fault-tolerance in general is a critical concept when designing and implementing high-availability systems.
High-availability systems are those that are designed to operate continuously without interruption and are essential for applications where downtime can have serious consequences, as in healthcare, finance, transportation, industry, information technology, and communication systems~\cite{gao2015survey,koren2020fault}.
Assuredly, fault-tolerance has many applications in various engineering fields such as automotive, aerospace, and avionics~\cite{baleani2003fault, yin2016review, edwards2010fault}, as well as distributed computing such as cloud computing~\cite{bala2012fault,amin2015review,jhawar2017fault,kumari2021survey}, and other distributed systems~\cite{cristian1991understanding,avizienis2004basic}.

Fault-tolerant distributed systems are attracting increasing interest due to the possibilities connected with the applications of blockchains particularly in cyber-physical systems such as IoT and Industry~4.0 applications. In addition, blockchain technologies are also used in finance, and other distributed computing applications such as Smart Contracts. Blockchain is considered a subgroup of distributed ledger technology (DLT). DLT requires a consensus protocol to commit transactions (e.g. read/write operations on the local storage of its members, or to perform some actuation scheme).
Indeed, blockchains are popular because they record information in a way that makes it difficult or impossible to change, hack, or cheat the system. Transactions are duplicated and distributed across the entire network of computer systems on the blockchain. However, data theft has seen an increasing threat, especially when financial transactions are concerned.

Blockchain technology is being integrated into IoT systems to improve their availability \cite{panarello2018blockchain}. A key benefit of blockchain technology is its ability to provide a decentralized infrastructure that can be used to manage and distribute resources across a network of devices. In IoT systems, this can be particularly useful for managing and distributing computational resources, such as processing power and storage capacity, to ensure that the system remains available and responsive even when individual devices fail \cite{boudguiga2017towards}. For example, blockchain-based smart contracts can be used to automatically allocate computational resources to IoT devices based on their needs and availability \cite{ozyilmaz2019designing}. Additionally, blockchain can be used to create a distributed ledger of device availability and resource usage, making it easier to manage and monitor the health of an IoT system.

Various techniques have been developed and implemented to enhance the fault tolerance of systems, aiming to ensure different levels of resilience against faults. This study specifically focuses on distributed computer systems that are capable of tolerating Byzantine faults ~\cite{lamport1982byzantine}. A Byzantine fault refers to a scenario where a node within the network behaves maliciously, such as sending conflicting messages to different servers or becoming unresponsive. It is important to note that this definition is distinct from a crash, which occurs when a node is not malicious but becomes unresponsive due to technical failures like power or connectivity outages. In both cases, whether it is a Byzantine fault or a crash, the system may encounter difficulties in achieving a consensus.

To implement a Byzantine Fault-Tolerant (BFT) consensus protocol, the minimum number of servers required is $N\geq4$ when the servers exchange unsigned messages.
Indeed, with unsigned messages, if $N<4$, the problem does not have a solution, as explained in the original article~\cite{lamport1982byzantine}.
In BFT systems, the term "quorum" refers to the minimum number of commit messages required to achieve consensus. A quorum is obtained when the count of responsive nodes that are honest, denoted as $h$, reaches a certain threshold defined as:
\begin{equation}\label{eq:system_available}
   h>2N/3.
\end{equation}
Hence, a BFT distributed system, in which unsigned messages are exchanged, can handle up to $f$ Byzantine faults, where
\begin{equation}
    f<N/3.
\end{equation}

Prior to the implementation of a BFT protocol, conducting an availability evaluation is an essential analysis that must be performed. This evaluation plays a vital role in ensuring the successful application of the system.
Indeed, the implementation of a computer network can be both costly and technically demanding, often leading to potential underperformance in terms of expected availability levels. In light of this consideration, analytical techniques to study a desired system have been developed.
The advantage of the analytical approach resides in the possibility of tuning the parameters characterizing the modelled system in a straightforward and inexpensive way.
This is particularly true for what concerns the development of DLTs, where analytical approaches applied to the study of network availability can address security bottlenecks caused by the malevolent nodes or crashes in the system.
This is critical particularly in a decentralized environment, because there is no central authority enforcing network policies or scheduling repairs, therefore a thoughtful knowledge of critical scenarios is necessary to overcome possible service downtime.
An evident downside is the difficulty of the development of an analytical model suitable to describe the network.
Nevertheless, over the past few decades, certain analytical methodologies such as continuous-time Markov chains (CTMC) have been extensively and effectively utilized to assess the availability of intricate systems.~\cite{goyal1987modeling,bolch2006queueing}.

In this article, we present an analytical model (based on CTMCs), that is able to evaluate BFT systems availability.
This model is employed to investigate the relationship between different parameters, in order to identify the best configurations to maximise the availability.
Since the occurrence of Byzantine nodes is not treated as a dynamic process, i.e. a malicious server does not change its stance over time, it is vital to understand what scenarios are to be expected in this framework.
A critical downside for this kind of assumption is the complex estimation of how many Byzantine nodes may be present in the network for a given configuration.
In~\cite{marcozzi2023availability}, the assumption of a degenerate distribution for the number of Byzantine servers in the system led to the definition of three levels of Byzantine threats; low, medium, and high level.
This work, instead, presents a novel approach to estimate the number of Byzantine nodes in a given network. Considering the number of Byzantine nodes to be the result of a stochastic process, the probabilistic distribution used for the number of malicious nodes 
depends on very specific features characterizing the system under investigation.
Indeed, this study proposes a methodology to describe the system availability in the presence of $f$ Byzantine actors, where the distribution of $f$ is a choice of the decision-maker/investigator or a characteristic of the system itself.
The methodology is useful in case the distribution of the values of $f$ is known because the model should accurately predict and reproduce the behaviour of the system under investigation.
Conversely, if the statistical properties of the process describing the number of Byzantine nodes are not known, this methodology allows testing different distributions, in order to outline possible scenarios and design the implementation of the system according to the retrieved information.
The advantage of this approach is two-fold: it provides a light-weight framework to assess system availability, indicating the best and worst cases without the need to actually develop and implement the system. The contributions of this study can be summarised as follows:
\begin{itemize}
   \item In this study, we propose an analytical approach to model Byzantine servers in the presence of malicious nodes and failures.
 \item An iterative algorithm is also presented to calculate availability for various distributions and parameters (distribution dependent) of a number of Byzantine faults.
 \item The method presented is tested for Uniform, Poisson, Binomial and Degenerate distributions.
\end{itemize}


The rest of this study is organized as follows. In Section \ref{sec:related}, a comprehensive review of prior research on the analytical availability models is presented. Section \ref{sec:model} describes the availability model proposed, along with its underlying assumptions. The mathematical prerequisites to derive a solution for the model are investigated in Section \ref{sec:analysis}. Section \ref{sec:results} presents the obtained results from the study. Lastly, Section \ref{sec:conclusion} summarizes the findings, discusses potential applications, and outlines future advancements for the presented model.

\section{Related Work}\label{sec:related}
The definition of availability can vary depending on the specific context, but it is generally characterized as the system's ability to carry out its intended operation at any given moment.
In particular, availability might be viewed as a failure-free operation at any time $t$, in which case it takes the name of point or instantaneous availability.
Point availability, denoted as $A(t)$, is formally defined as the probability that the analyzed component functions correctly at a given time $t$.


\begin{equation}
A(t)=W(t)+\int_0^t W(t-x)\,r(x)\,dx \nonumber
\end{equation}

here $W(t)$ is the probability of not experiencing any failures for the component in the interval $(0, t]$ whereas $r(x)$ is the repair frequency.
Clearly, the equation demonstrates that the system achieves availability either when there are no failures within the interval $(0, t]$, or if failures do occur, they are promptly repaired prior to time $t$ \cite{trivedi2008probability}. However, availability can be also studied in a non-transient scenario, where the average probability to have the component functioning is considered.
Therefore, by employing the concepts of mean time to failure ($MTTF$) and mean time to repair ($MTTR$), it is possible to express the limiting availability $A$ as follows:

\begin{equation}
A=\lim_{t\longrightarrow\infty}A(t)=\frac{MTTF}{MTTF+MTTR}
\end{equation}
\medskip

It is important to note that the limiting availability is solely dependent to the values of $MTTF$ and $MTTR$, irrespective of the specific probability distributions governing the failure as well as repair times.

The assessment of availability in multi-server systems has gained significant importance in the relevant literature. Much information on the subject can be found in research articles, books, and reviews \cite{trivedi2017reliability}. Markov chain-based availability models are commonly employed in numerous studies, alongside other examples, to analyze and assess system availability.

The work in \cite{strielkina2018availability} presents an analytical approach to the availability of healthcare IoT infrastructures. The authors employ two-dimensional CTMCs to depict the availability of the considered IoT systems for healthcare infrastructure and its end-nodes. Additionally, they present an approach that uses Markov models to examine attacks targeting the potentially vulnerable aspects of healthcare IoT systems. The model includes a state diagram representing attacks on the IoT infrastructures in healthcare. The authors analyze the system's availability with respect to the flow intensities of service requests, emphasizing safety concerns and security-related issues in the healthcare IoT context.
Similarly in \cite{tang2021availability}, an investigation is conducted on the availability of healthcare IoT systems. The authors outline two categories of structures comprising the IoT system, utilizing separate two-dimensional Markov state-space models for the systems considered. They subsequently solve the equilibrium equations of the system employing a similar approach to the one described in the preceding article. The authors present various performance metrics related to availability, such as the probability of being able to provide service at full capacity,  reduced performance service, and the probability of having the system in a state which is not providing any services at all.
In addition, the work presented in \cite{rodrigues2021performance} showcases the performability evaluation of a smart hospital architecture to guarantee the quality of service in healthcare. The model employs two Stochastic Petri Nets, allowing the tuning of several parameters to adjust different scenarios and identify the most critical components of the architecture. The authors also present some results based on three possible scenarios, where a best-case result is obtained in the scenario where redundancy is implemented.

Several studies have focused on analyzing the availability of IoT systems, including \cite{pereira2021analytical} and \cite{ever2019performance}. Additionally, some studies such as \cite{kirsal2015modelling} have modelled the facilitating infrastructures in presence of failures. In \cite{pereira2021analytical}, the authors present analytical models to evaluate the availability considering various physical edge as well as  fog nodes used in  various applications. They compute $MTTF$ and $MTTR$ values for the systems considered and present a two-dimensional model that includes both failures and repairs. Similarly, in \cite{ever2019performance}, in addition to availability and  performance, the authors also evaluate the energy consumption-related measures of clustered IoT systems. They solve two-dimensional models for steady-state probabilities, which they use to compute crucial measures related to availability (e.g. the probability of being in a state where the system is fully operational) and assess other measures of performance such as the mean value of energy consumption.

Considering the modeling of cloud infrastructures, especially those based on Infrastructure as a Service (IaaS), scalability becomes a significant limiting factor. In the publication by Ataie et al. \cite{ataie2017hierarchical}, scalability challenges are addressed through the utilization of approximate Stochastic Reward Net (SRN) models, combined with folding and fixed-point iteration techniques. They use various failure and repair rates in their approach accurately capturing the characteristics of failures as well as repairs for physical machines in order to enable the analysis of availability-related functionalities.

The article by Longo et al. \cite{longo2011scalable} concentrates on attaining high availability in IaaS cloud systems. To expedite the analysis and resolution process, the authors employ an approach based on interacting Markov chains. They employ SRNs to compute the metrics of the Markov chains. The study includes a trade-off analysis between longer Mean Time to Failure ($MTTF$) and faster Mean Time to Repair ($MTTR$) in terms of system availability. Additionally, the impact of incorporating multiple concurrent facilities for repair facilities is examined.

In \cite{ever2017performability}, a novel approach with an approximate solution is introduced to address the potentially large numbers of servers in cloud based systems. The analytical models and solutions proposed in the study are comprehensive, yet they are capable of handling substantial numbers of nodes, ranging from hundreds to thousands. The research focuses on the quality of service provided by cloud centers, taking into account both server availability and performability metrics considering server failures as well as repairs. Notably, this study distinguishes itself from other reviewed works by emphasizing the ability to analyze and model large-scale cloud systems while incorporating considerations for server availability and quality of service.

In \cite{melo2021distributed}, the focus is on blockchain-based systems that can provide service over cloud infrastructures. The research introduces models to assess the availability and capacity-oriented availability of cloud computing infrastructures that host distributed applications using the Ethereum blockchain platform. The conventional approach is employed to represent the system's availability by considering the ratio of $MTTF$ to $MTTR$. The availability outcomes are depicted as functions of $MTTF$ and $MTTR$ for servers as well as miner and bootnodes.

The aforementioned studies employ analytical models to evaluate the availability of various distributed systems.
Similar to these studies, in this study as well the approach presented assumes that the time between failures and repair times adheres to an exponential distribution.
The main focus of this article is to analyze the availability of a system through the utilization of a Markov chain-based model. When the existing studies in the field are investigated, we see that many studies have employed Markov processes as a common formalism and terminology for modeling various systems. However, based on the authors' knowledge, they are the first to utilize a Markov formalism specifically for modeling the availability of BFT systems, as previously introduced in a previous work~\cite{marcozzi2023availability}.
By introducing a Markov-based approach, the authors aim to provide a novel perspective and contribution to the analysis of BFT system availability. This approach allows for a systematic examination of the system's behavior and performance in terms of availability, considering the unique characteristics and challenges associated with BFT systems.
The utilization of a Markov formalism in the context of BFT system availability modeling distinguishes this work from previous studies and emphasizes its novelty and potential impact in understanding and evaluating the availability of such systems.

\section{System Model}\label{sec:model}
\begin{figure*}[!h]
    \centering
    \includegraphics[height=0.33\textheight]{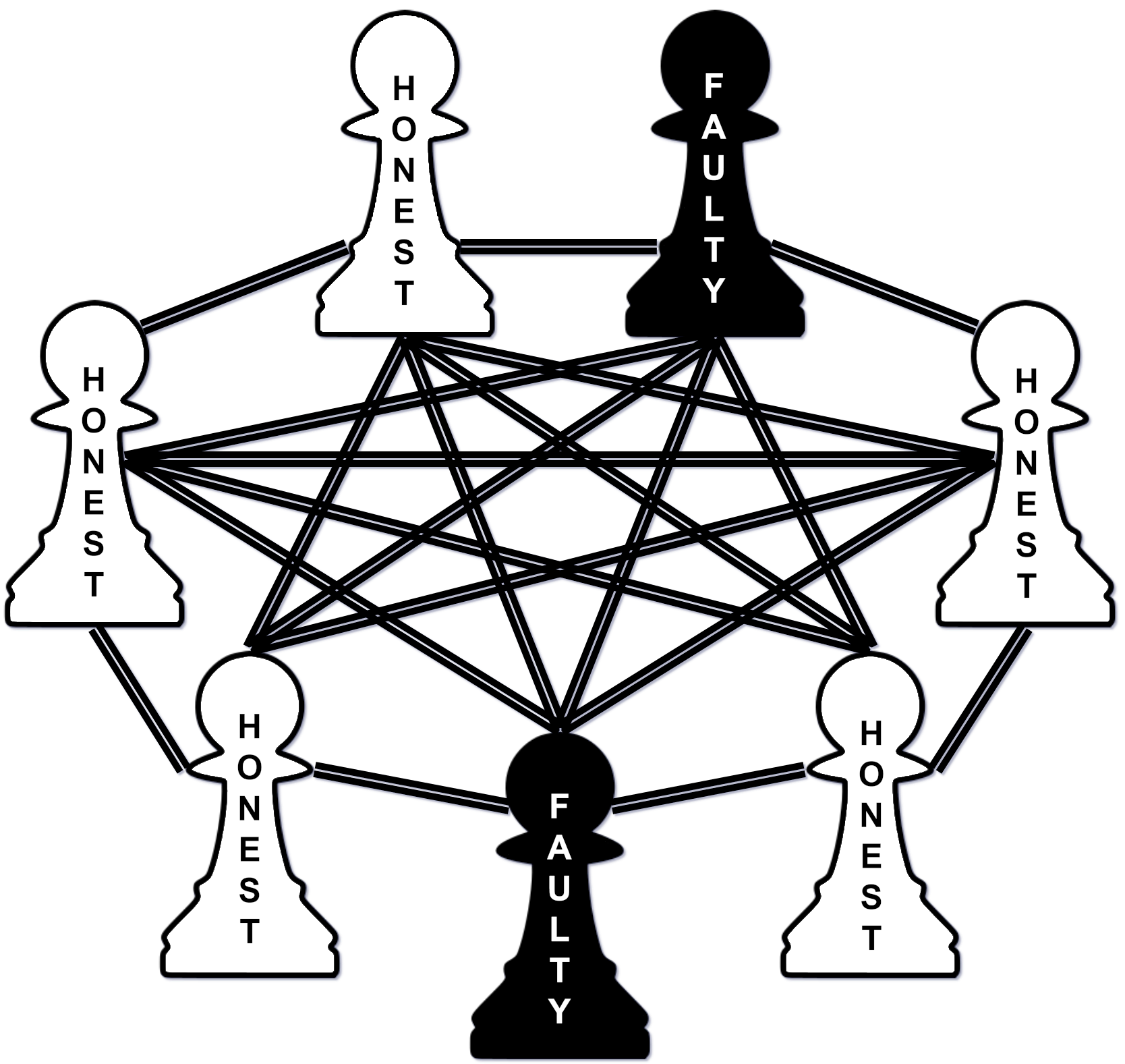}
    \caption{Byzantine Fault-Tolerant (BFT) system $N=7$.}
    \label{fig:BFT_scheme}
\end{figure*}

The system is designed as a network comprising $N$ nodes, with the responsibility of collaborating to get an agreement on specific tasks. These nodes aim to reach a consensus on those tasks through message exchange among themselves. The communication can occur in any way suitable for the applications, such as end-to-end, as shown in \autoref{fig:BFT_scheme}. In the diagram, the white pawns symbolize non-Byzantine nodes, while the black pawns represent Byzantine nodes.

In this study, the proposed scheme is used to analyze a system with $N$ number of servers. These servers are also referred to as replicas or nodes with the main duty of committing messages to their storage. This scenario is commonly seen in distributed databases or blockchain networks. Since the scheme is applicable to various systems, we use this particular example as a practical application.

The availability model being used is a quasi-birth-death process based on a CTMC. In this type of stochastic process, the random variables have an exponential distribution, and the system can switch between states at rates specified by the stochastic transition matrix. The Markov property is satisfied, meaning that the distribution of the state probabilities for future states depends solely on the present state and not on past states.

In this model, the parameters $\xi$ and $\eta$ represent the breakdown and repair  rates respectively, as illustrated in \autoref{fig:availability_model}.
In this model, server breakdowns are independent. In an event of a failure, the broken servers are considered for repair one at a time. Consequently, the breakdown rate $\xi$ is scaled considering the number of available nodes $f$. In other words, $f\xi$ is the break-down rate for the state where there are $f$ nodes, and $\xi$ is the corresponding break-down rate when only one node is available. However, the repair process can only occur for one node assuming a single repair facility with a repair rate of $\eta$.

\input{img/availability_figure}

The model \autoref{fig:availability_model} is proposed with the following assumptions\footnote{To facilitate clarity in presentation, we assume that $N, h, f\in\mathbb{N}_0$. Consequently, when performing divisions, the \textit{ceiling} $\lceil\cdot\rceil$ and \textit{floor} $\lfloor\cdot\rfloor$ functions are implicitly applied as appropriate.}: there are $N$ servers in the system. $h\leq N$ of these nodes are honest nodes taking part in the network operations as expected. $f\leq N$ of the nodes are malevolent nodes.
In this context, $H:\{h\in\mathbb{N}_0\,|\,h\leq N]\}\rightarrow\{h\in\mathbb{N}_0\,|\,h\leq N]\}$ and $F:\{f\in\mathbb{N}_0\,|\,f\leq N]\}\rightarrow\{f\in\mathbb{N}_0\,|\,f\leq N]\}$ can be treated as random variables following an arbitrary distribution.
$H$ and $F$ are dependent on each other, such that their realizations sum up to $N$, i.e. $H(h)+F(f)=h+f=N$.
Therefore, since $N$ is considered to be a constant, $H$ and $F$ are dependent random variables and their outcomes can be written as $f=N-h$ or $h=N-f$, considering either $H$ or $F$ to be the independent random variable.
Without loss of generality, $F$ is the independent discrete random variable. Hence, $h=N-f$ is a realization dependent on the value of the discrete random variable $F$.
The reasons behind this abstraction are presented in \autoref{sec:introduction}.

The state diagram shown in \autoref{fig:availability_model} consists of $(h+1)(f+1)$ states. It is worth noting that every state in the chain can be reached from any initial state, demonstrating the chain's property of being both irreducible and ergodic. These two conditions are adequate for the chain to possess a stationary distribution. This means that, given a system with specific parameters, it is possible to compute the system's limiting availability using the stationary probability distribution associated with its CTMC.
To calculate the stationary distribution of the chain, it is necessary to formulate the generating equations as a system of linear equations where the state probabilities for given transition rates can be determined.
Such a system of simultaneous equations is known as Kolmogorov equations.
The following equation represents the whole system of linear equations:
\begin{multline}\label{eq:balance}
    \left[(2-\delta_{ih}-\delta_{jf})\eta + (i+j)\xi\right] P_{i,j} + \\
    - \eta \left[P_{i,j-1}(1-\delta_{j0}) + P_{i-1,j}(1-\delta_{i0})\right] + \\
    - (i+1)\xi P_{i+1,j}(1-\delta_{ih}) - (j+1)\xi P_{i,j+1}(1-\delta_{jf}) = 0,
\end{multline}
where $\delta_{ij}$ indicates the Kronecker delta, i.e. $\delta_{ij}=1$ if $i=j$ and $\delta_{ij}=0$ if $i\neq j$.
In a compact form, \autoref{eq:balance} describes all the possible equations in the system by varying the indices $i$ and $j$, where $i\in[0,h]$ and $j\in[0,f]$.
Thus, consistently with number of possible states, there are $(h+1)(f+1)$ equations to be solved simultaneously.
However, because the elements of $\Vec{P}$, $P_{i,j}$, are probabilities, the additional condition $\sum_i\sum_j P_{i,j}=1$ is imposed.

In \autoref{eq:balance}, $P_{i,j}$ is the probability that the system is in state $(i,j)$, while the coefficients of $P_{i,j}$s are the entries in a coefficient matrix, $\mathbf{Q}$.
Indeed, $\mathbf{Q}$ represents the stochastic transition matrix for CTMC under consideration, encompassing the transition rates from one state to another.

Note that, mirroring the lattice structure of the model, it seems natural to write the elements of $\Vec{P}$ using the indices $i$ and $j$, although $\dim\mathbf{Q}=(h+1)(f+1)\times(h+1)(f+1)$, because there are $(h+1)(f+1)$ states in the system.
This means that it is not proper to use $i$ and $j$ while computing the elements of $\Vec{P}$.
Indeed, $\Vec{P}$ is having a matrix structure when expressed as $P_{i,j}$, therefore it has to be flattened into a vector with elements $P_{i}$, where $i\in\left[0,(h+1)(f+1)\right]$.
This last remark is important, because it ensures that the dimensions of $\Vec{P}$ and $\mathbf{Q}$ are matching, since the matrix of coefficients $\mathbf{Q}$ has indices $i,j\in\left[0,(h+1)(f+1)\right]$.

The system of simultaneous equations derived from Equation \ref{eq:balance} can be expressed straightforwardly in the form $\mathbf{Q}\Vec{P}=0$, where $\mathbf{Q}$ represents the coefficient matrix, $\Vec{P}$ denotes the vector of unknowns, and $0$ corresponds to the vector of constants (zero). To solve homogeneous matrix equations there are a plethora of techniques, even though some of them might not be applied in this context.
For instance, since $\mathbf{Q}$ is, indeed, a singular matrix, commonly used methods of linear algebra to solve matrix equation, such as LU decomposition or Gauss elimination, cannot be applied for singular matrices. It is assumed that a non-trivial solution to the system of simultaneous linear equations exists. Notably, matrix $\mathbf{Q}$ possesses at least one singular value equal to zero, indicating the presence of a non-trivial solution to the linear simultaneous equations. To address this, the Singular Value Decomposition (SVD) method is employed.

\section{Availability Analysis}\label{sec:analysis}
In~\cite{marcozzi2023availability}, a stringent premise regarding the occurrence of Byzantine nodes is adopted.
Specifically, it is assumed that the threat level due to Byzantine nodes is either low, medium, or high, with a different number $f$ for each of the three levels.
While this is a pragmatic assumption, it may not, in principle, reflect the statistics of a real implementation.
From this observation, it is clear that, the analysis of the availability may be influenced by the method used to describe the occurrences of Byzantine nodes.
Therefore, a new paradigm is presented, in which the number of Byzantine nodes $f$ is determined by the random variable $F$.

Note that the study of the stochastic properties associated with the distribution of the number of Byzantine nodes in the network is not affecting the analytical model used to describe the system.
In other words, a separate layer of abstraction is added on top of the availability model, in order to analyse the system in a more general way.
This additional abstraction can be considered as an experimental framework, in which the experimenter/decision-maker is testing the system.
Given the system parameters ($N,\eta,\xi$) and a probability distribution $Pr(F=f)=p(f)$, the methodology to observe is composed by the steps reported in Algorithm~\ref{alg:mean_availability}.
\begin{algorithm}
\caption{Pseudo-code to calculate availability with $f$ as random variable}\label{alg:mean_availability}
\begin{algorithmic}
\Require $N,N_{max} \geq 4$ and $\xi,\eta>0$ and $\xi/\eta \ll 1$
\For{$N\leq N_{max}$}
\State $f \gets 0$
\While{$f<N/3$}
\State $h \gets N-f$
\State $\mathbf{Q} \gets \mathbf{Q}(N,f,h,\xi,\eta)$
\State $\Vec{P} \gets SVD(\mathbf{Q},0)$ \Comment{compute state probabilities through SVD}
\State $A_{h,f} \gets \sum_{i>2N/3}^{h}\sum_{j=0}^{f}P_{i,j}$ \Comment{availability}
\State $f\gets f+1$
\EndWhile
\State $\Bar{A} \gets \sum_{h,f} p(f)\,A_{h,f}$ \Comment{mean availability}
\EndFor
\end{algorithmic}
\end{algorithm}

The process described in Algorithm~\ref{alg:mean_availability} requires a maximum number of nodes to be considered $N_{max}\geq4$ and rates $\xi,\eta>0$ such that $\xi/\eta\ll1$.
The process starts setting the value $f=0$, hence imposing $h=N-f$.
The matrix $\mathbf{Q}$ is determined using \autoref{eq:balance}, thus $\Vec{P}$ can be computed through Singular Value Decomposition.
For the pair ($h$, $f$), the probabilities $P_{i,j}$ are computed and in turn arranged in matrix $\mathbf{P}$ (vector $\Vec{P}$ is transformed into matrix $\mathbf{P}$ to align with the two-dimensional structure depicted in Figure \ref{fig:availability_model}.).
Finally, for the resulting set ($N,f,h,\eta,\xi$), availability can be calculated.
Availability refers to the cumulative probability that the system is operational and capable of committing messages.
As prescribed in \autoref{eq:system_available}, the system is available for all the states with $i>2N/3$, consequently, the corresponding state probabilities are aggregated to calculate the availability.:
\begin{equation}\label{eq:availability}
    A_{h,f} = \sum_{i>\frac{2N}{3}}^{h}\sum_{j=0}^{f}P_{i,j}.
\end{equation}
At each iteration, $f$ is increased by $1$.
The algorithm iterates until a predefined value of $N_{max}$ is reached.
After the iterative part, there is a resulting collection of $A_{h,f}$s, one for each ($N,f,h,\eta,\xi$).
Therefore, the mean value of the availability is
\begin{equation}
    \Bar{A}=\sum_{h,f} p(f)\,A_{h,f}.
\end{equation}
Essentially, the procedure described above compute the mean availability for a system with $N$ servers (subjected to break-down and repair processes at rate $\xi$ and $\eta$), where the number of Byzantine actors in the system, $f$, is deriving from the realizations of a random variable $F$ distributed according to an arbitrary probability distribution (see \autoref{tab:probabiliy_distributions} for a concise description of the probability distributions used to determine $f$). This means that the procedure in Algorithm~\ref{alg:mean_availability} can be iterated over a range of several $N$ and different probability distributions for $F$.
In this way, the behaviour of the system's availability, for different distributions, can be studied as a function of the number of servers.
Similarly, to study the relationship between rates $\xi,\eta$ and availability, the probability distribution for $Pr(F=f)$ can be fixed and then it can be computed the availability of the system at the variation of $\xi$ and $\eta$, for different $N$.

A special attention should be reserved to the analysis of the Poisson distribution. The pmf of Poisson distribution is defined on the positive integers, therefore a truncated version of the pmf is needed to match the domain of definition $[0,N]$ for the occurrence of Byzantine nodes. The right-truncated Poisson distribution~\cite{johnson2005univariate} is defined as
\begin{equation*}
    p(x;\lambda, N) = \begin{cases}\frac{\lambda^x}{x!}\left(\sum_{y=0}^{N}\frac{\lambda^y}{y!}\right)^{-1}, & x=0,1,\ldots,N\\
    0, & \text{otherwise}\end{cases}
\end{equation*}
that is derived from the definition of Poisson distribution, in which the series representation of the exponential is truncated to $N$.
The mean can be computed from the definition of expected value $\mu=E[X]=\sum_{x=0}^{N}x\,p(x)=\lambda\frac{N\,\Gamma(N,\lambda)}{\Gamma(N+1,\lambda)}$, where $\Gamma(N,\lambda)$ is the incomplete gamma function.

\renewcommand{\arraystretch}{2}
\begin{table}[!h]
    \centering
    \begin{tabular}{llcc}
       \textbf{Distribution} & \textbf{pmf} & \textbf{Mean $\mu$} & \textbf{Variance $\sigma^2$} \\
       \hline
        Uniform & $p(x;a,b)={\frac {1}{b-a+1}}$ & $\frac{b+a}{2}$ & $\frac{(b-a+1)^2-1}{12}$  \\
        \makecell[lc]{Right-truncated\\Poisson} & $p(x;\lambda, n)=\frac{\lambda^x}{x!}\left(\sum_{y=0}^{n}\frac{\lambda^y}{y!}\right)^{-1}$ & $\lambda\frac{n\,\Gamma(n,\lambda)}{\Gamma(n+1,\lambda)}$ & - \\
        Binomial & $p(x;n,q)=\binom{n}{x}q^x(1-q)^{n-x}$ & $nq$ & $nq(1-q)$ \\
        Degenerate & $p(x;x_0) = \delta_{x\,x_0}$ & $x_0$ & $0$
    \end{tabular}
    \caption{A summary of the probability distributions used in this work to characterize the occurrences of Byzantine nodes, with $N\in\left[4,128\right]$. pmf indicates probability mass function, $\mu$ the mean of each distribution, and $\sigma^2$ the variance of the distribution.
    Parameters for each distributions are specified in the next section, in correspondence of the two comparative results: \autoref{fig:comparison} and \autoref{fig:comparison_middle}.
    }
    \label{tab:probabiliy_distributions}
\end{table}

Regarding the proposed methodology, note that it is vital to choose appropriately the parameters of the arbitrary probability function generating the random values $f$.
While it is out of the scope for this study to determine whether there is an \textit{a priori} restriction on which probability function to use in characterizing the occurrences of Byzantine nodes, it is advisable to properly select the first two moments, i.e. mean and variance, of any chosen distribution.
To better explain this, consider the impact that the parameters of the probability distribution have: if the mean $\mu$ is outside the interval $[0,N/3)$ and the probability function is narrow (low variance), several zero-valued availability numbers will be sampled; same situation would occur if $\mu\in[0,N/3)$, but the variance is high; an optimal choice, instead, is represented by the distribution not spreading excessively and $\mu\in[0,N/3)$.

Lastly, this methodology recreates the results presented in our previous study~\cite{marcozzi2023availability}, where a constant number $f$ is selected to reflect a threat level due to the ratio of Byzantine nodes in the system.
This validates the observation that, when defining some possible threat levels of the system, the decision-maker is, indeed, assuming a degenerate distribution for $F$, i.e. a constant value $f$ representing the number of Byzantine nodes in a system of $N$ nodes.

\section{Results and Discussions}\label{sec:results}

In this section, results are provided to show the effects of various distributions of the Byzantine faults on availability.

\autoref{fig:comparison}, shows the effects of four different probability distributions for the random variable $F$ (as from \autoref{tab:probabiliy_distributions}), on mean system availability for $N\in[4,128]$ and $\xi/\eta=0.015$. Different lines represent a separate choice of a probability distribution for the value of $f$. Parameters of the uniform distribution are $a=0$ and $b=N$. $\lambda=N/6$ is used for the right-truncated Poisson distribution, in which $\Gamma(n,\lambda)$ is the incomplete gamma function. For the binomial distribution $n=N$ and $q=1/6$, where $\binom{n}{x}$ is the binomial coefficient. Lastly, the degenerate distribution uses the Kronecker delta $\delta_{x\,x_0}$, with $x_0=N/6$. In the figure, the uniform distribution has the mean $\mu=N/2$, while all the other distributions have the mean $\mu=N/6$, the center of the interval $[0,N/3)$.
The figure shows that, for different choices of the probability distribution of $F$, there is a distinctive behaviour of the mean availability. This behaviour varies between the worst-case scenario, which can be observed when the random variable $F$ is drawn from a uniform distribution to the best case, where a degenerate distribution with constant value $f=N/6$ is used.
However, this configuration for the uniform distribution is expected to give the worst-case scenario, since the mean of the distribution is centered around the middle of the interval for the values of $N$, while the other distributions are centered around $N/6$.

\begin{figure}[!h]
    \centering
    \begin{tikzpicture}
\begin{axis}[title = {Mean availability @ $\xi/\eta=0.015$}, width=.63\textwidth,
            xlabel = {$N$}, ylabel = {$A$},
            xmin = 4, xmax = 128,
            ymin = 0, ymax = 1,
            mark size=1.pt,
            cycle list name=exotic,
            legend style={nodes={scale=0.6, transform shape}},
            legend pos=south west]
\addlegendimage{empty legend}
\addlegendentry{\hspace{-.6cm}\textbf{Distribution}}
\addplot+[smooth
    ] table [x index = 0, y index = 1, col sep=comma]{data/distribution_comparison.csv};
    \addlegendentry{uniform}
\addplot+[smooth
    ] table [x index = 0, y index = 2, col sep=comma]{data/distribution_comparison.csv};
    \addlegendentry{Poisson}
\addplot+[smooth
    ] table [x index = 0, y index = 3, col sep=comma]{data/distribution_comparison.csv};
    \addlegendentry{binomial}
\addplot+[smooth
    ] table [x index = 0, y index = 4, col sep=comma]{data/distribution_comparison.csv};
    \addlegendentry{degenerate}
\end{axis}
\end{tikzpicture}
    \caption{Availability as a function of the number of servers and fixed ratio $\xi/\eta$.}
    \label{fig:comparison}
\end{figure}

\begin{figure}[!h]
    \centering
    \begin{tikzpicture}
\begin{axis}[title = {Mean availability @ $\xi/\eta=0.015$ and $\mu=N/2$}, width=.62\textwidth,
            xlabel = {$N$}, ylabel = {$A$},
            xmin = 4, xmax = 128,
            ymin = 0, ymax = 0.42,
            mark size=1.pt,
            cycle list name=exotic,
             legend style={nodes={scale=0.7, transform shape}},
            legend pos=north east]
\addlegendimage{empty legend}
\addlegendentry{\hspace{-.6cm}\textbf{Distribution}}
\addplot+[smooth
    ] table [x index = 0, y index = 1, col sep=comma]{data/distribution_comparison_middle.csv};
    \addlegendentry{uniform}
\addplot+[smooth
    ] table [x index = 0, y index = 2, col sep=comma]{data/distribution_comparison_middle.csv};
    \addlegendentry{Poisson}
\addplot+[smooth
    ] table [x index = 0, y index = 3, col sep=comma]{data/distribution_comparison_middle.csv};
    \addlegendentry{binomial}
\end{axis}
\end{tikzpicture}
    \caption{The variation in system availability as a function of the number of servers and fixed ratio $\xi/\eta$.}
    \label{fig:comparison_middle}
\end{figure}

\autoref{fig:comparison_middle} presents the behaviour of the mean system availability for $N\in[4,128]$ and $\xi/\eta=0.015$, when the mean of each probability distribution is $\mu=N/2$. In this figure, different lines represent a separate choice of probability distribution for the value of $f$. Parameters of the uniform distribution are $a=0$ and $b=N$. For the right-truncated Poisson distribution, in which $\Gamma(n,\lambda)$ is the incomplete gamma function, $\lambda=N/2$ is used. The binomial distribution has $n=N$ and $q=1/2$, where $\binom{n}{x}$ is the binomial coefficient. As expected, the degenerate distribution, when $f=N/2$, gives availability that is constantly zero, therefore it is not reported. In this graph, the best case is the one in which the uniform distribution is employed, while the worst case occurs when the binomial distribution describes the occurrence of Byzantine nodes in the system.
Differently from \autoref{fig:comparison}, with this configuration, the uniform distribution is clearly the distribution giving the best result in \autoref{fig:comparison_middle}. This is because the probability to get a value $f<N/3$, such that the quorum is reached, is higher for the uniform distribution than for the other distributions. This is simply because, while the mean is the same for the selected distributions, the variance of the possible values of $f$ is larger for the uniform distribution, hence there is a higher probability to select a value $f$ satisfying the quorum.

\autoref{fig:different_xi} presents an example of system availability trend for different ratios of $\xi/\eta$.
Here $F$ is distributed according to the degenerate distribution centered around the value $f=N/6$. The plot shows how the availability of the system is degrading when the ratio $\xi/\eta$ is increasing.

\begin{figure}[!h]
    \centering
    \begin{tikzpicture}
\begin{axis}[title = {Average availability (degenerate distribution with $f=N/6$)}, width=.69\textwidth,
            xlabel = {$N$}, ylabel = {$A$},
            xmin = 4, xmax = 128,
            ymin = 0, ymax = 1,
            mark size=1.pt,
            legend style={nodes={scale=0.7, transform shape}},
            legend pos=south west]
\addlegendimage{empty legend}
\addlegendentry{\hspace{-.6cm}\textbf{$\xi/\eta$}}
\addplot+[smooth
    ] table [x index = 0, y index = 1, col sep=comma]{data/xi_example.csv};
    \addlegendentry{$0.01$}
\addplot+[smooth
    ] table [x index = 0, y index = 2, col sep=comma]{data/xi_example.csv};
    \addlegendentry{$0.015$}
\addplot+[smooth
    ] table [x index = 0, y index = 3, col sep=comma]{data/xi_example.csv};
    \addlegendentry{$0.02$}
\end{axis}
\end{tikzpicture}
    \caption{System availability as a function of the number of servers with ratio $\xi/\eta=0.01,0.015,0.02$.}
    \label{fig:different_xi}
\end{figure}

Please note that the values of availability are not represented by a smooth line because some numbers for $N$ correspond to optimal configurations of BFT systems.
For instance, any $N$ satisfying the equation $(N \mod 3) = 1$, $N\geq4$, produces a system with better availability than the ones generated by $N-1$ and $N-2$, e.g., the value of availability when $N=16$ is higher than when $N=15$ or $N=14$.

In summary, from this study, it can be concluded that system availability is indeed non-linearly dependent on the number of the servers in the network. This relation is inversely proportional to the number of the servers. Moreover, results show that the occurrence of Byzantine nodes in the system effects the overall availability, especially with regards to the probability distribution describing this phenomenon and its parameters. Finally, the ratio between break-down rate and repair rate, likewise, regulates the value of availability for the system, with lower values at the increase of the ratio $\xi/\eta$.

\section{Conclusion and future work}\label{sec:conclusion}

Distributed systems are widely used in various engineering sectors, industrial production, data analysis and management, cryptocurrencies, and more. Ensuring fault tolerance and security against malicious attacks has become increasingly important, particularly in scenarios where high availability is crucial. The uninterrupted operation of vital applications requires a well-designed distributed system capable of handling various potential scenarios.

BFT protocols play a significant role in achieving fault tolerance. These protocols were developed to model consistent distributed computer networks and parallel computing. BFT systems can maintain resilience even when malicious actors are involved in pursuing a common goal. Computer networks, including DLTs, like blockchains, often employ BFT algorithms to ensure continuous system operation.

we present an analytical availability model designed to assess fault-tolerant multi-server systems. The model leverages CTMCs to analyze the availability of Byzantine Fault-Tolerant (BFT) systems, taking into account breakdowns, repairs, and the presence of malicious nodes. The analysis considers a range of total nodes, $N$, from 4 to 128, and incorporates various probability distributions to model the proportion of malicious nodes. The numerical results exhibit the relationship between availability and the number of participants, as well as the relative number of honest actors, utilizing various probability distributions that represent the number of malicious nodes.

This work makes a significant contribution by expanding the availability modeling to incorporate the existence of malicious nodes with arbitrary non-deterministic probabilistic distributions. The model unveils a non-linear association between the number of servers and availability, where availability is inversely related to the number of nodes in the system, regardless of the distribution tested. This relationship becomes stronger as the ratio of breakdown rate to repair rate increases. Furthermore, the model serves as an initial step in performability modeling of distributed systems based on BFT consensus protocols.



 \bibliographystyle{elsarticle-num} 
 \bibliography{bibliography}
 
\end{document}

%% file: img/availability_figure.tex
\begin{figure}[!h]
	\centering
        \resizebox{!}{.85\textwidth}{%
	\begin{tikzpicture}[->, >=stealth', auto, semithick, node distance=3.5cm, transform shape]
	\tikzstyle{every state}=[fill=white,draw=black,thick,text=black,scale=.9]
				
				\node[state, minimum size=1.6cm] (0 0) {$0,0$};
				
				\node[state, draw=none, minimum size=1.6cm] (none00) [right of=0 0] {$\ldots$};
				\path (0 0) edge[bend right,below] node{$\eta$} (none00);
				\path (none00) edge[bend right,above] node{$\xi$} (0 0);
				
				\node[state, minimum size=1.6cm] (0 f-j) [right of=none00] {$0,j$};
				\path (none00) edge[bend right,below] node{$\eta$} (0 f-j);
				\path (0 f-j) edge[bend right,above] node{$j\xi$} (none00);
				
				\node[state, draw=none, minimum size=1.6cm] (none02) [right of=0 f-j] {$\ldots$};
				\path (0 f-j) edge[bend right,below] node{$\eta$} (none02);
				\path (none02) edge[bend right,above] node{$(j+1)\xi$} (0 f-j);
				
				\node[state, minimum size=1.6cm] (0 f) [right of=none02] {$0,f$};
				\path (none02) edge[bend right,below] node{$\eta$} (0 f);
				\path (0 f) edge[bend right,above] node{$f\xi$} (none02);
				
				\node[state, draw=none, minimum size=1.6cm] (none10) [below of=0 0] {$\vdots$};
				\path (0 0) edge[bend right,left] node{$\eta$} (none10);
				\path (none10) edge[bend right,right] node{$\xi$} (0 0);
				
				\node[state, draw=none, minimum size=1.6cm] (none11) [below of=0 f-j] {$\vdots$};
				\path (0 f-j) edge[bend right,left] node{$\eta$} (none11);
				\path (none11) edge[bend right,right] node{$\xi$} (0 f-j);
				
				\node[state, draw=none, minimum size=1.6cm] (none12) [below of=0 f] {$\vdots$};
				\path (0 f) edge[bend right,left] node{$\eta$} (none12);
				\path (none12) edge[bend right,right] node{$\xi$} (0 f);
				
				\node[state, minimum size=1.6cm] (h-i 0) [below of=none10] {$i,0$};
				\path (none10) edge[bend right,left] node{$\eta$} (h-i 0);
				\path (h-i 0) edge[bend right,right] node{$i\xi$} (none10);
				
				\node[state, draw=none, minimum size=1.6cm] (none20) [right of=h-i 0] {$\ldots$};
				\path (h-i 0) edge[bend right,below] node{$\eta$} (none20);
				\path (none20) edge[bend right,above] node{$\xi$} (h-i 0);
				
				\node[state, minimum size=1.6cm] (h-i f-j) [right of=none20] {$i,j$};
				\path (none20) edge[bend right,below] node{$\eta$} (h-i f-j);
				\path (h-i f-j) edge[bend right,above] node{$j\xi$} (none20);
				\path (none11) edge[bend right,left] node{$\eta$} (h-i f-j);
				\path (h-i f-j) edge[bend right,right] node{$i\xi$} (none11);
				
				\node[state, draw=none, minimum size=1.6cm] (none22) [right of=h-i f-j] {$\ldots$};
				\path (h-i f-j) edge[bend right,below] node{$\eta$} (none22);
				\path (none22) edge[bend right,above] node{$(j+1)\xi$} (h-i f-j);
				
				\node[state, minimum size=1.6cm] (h-i f) [right of=none22] {$i,f$};
				\path (none22) edge[bend right,below] node{$\eta$} (h-i f);
				\path (h-i f) edge[bend right,above] node{$f\xi$} (none22);
				\path (none12) edge[bend right,left] node{$\eta$} (h-i f);
				\path (h-i f) edge[bend right,right] node{$i\xi$} (none12);
				
				\node[state, draw=none, minimum size=1.6cm] (none30) [below of=h-i 0] {$\vdots$};
				\path (h-i 0) edge[bend right,left] node{$\eta$} (none30);
				\path (none30) edge[bend right,right] node{$(i+1)\xi$} (h-i 0);
				
				\node[state, draw=none, minimum size=1.6cm] (none31) [below of=h-i f-j] {$\vdots$};
				\path (h-i f-j) edge[bend right,left] node{$\eta$} (none31);
				\path (none31) edge[bend right,right] node{$(i+1)\xi$} (h-i f-j);
				
				\node[state, draw=none, minimum size=1.6cm] (none32) [below of=h-i f] {$\vdots$};
				\path (h-i f) edge[bend right,left] node{$\eta$} (none32);
				\path (none32) edge[bend right,right] node{$(i+1)\xi$} (h-i f);
				
				\node[state, minimum size=1.6cm] (h 0) [below of=none30] {$h,0$};
				\path (none30) edge[bend right,left] node{$\eta$} (h 0);
				\path (h 0) edge[bend right,right] node{$h\xi$} (none30);
				
				\node[state, draw=none, minimum size=1.6cm] (none40) [right of=h 0] {$\ldots$};
				\path (h 0) edge[bend right,below] node{$\eta$} (none40);
				\path (none40) edge[bend right,above] node{$\xi$} (h 0);
				
				\node[state, minimum size=1.6cm] (h f-j) [right of=none40] {$h,j$};
				\path (none40) edge[bend right,below] node{$\eta$} (h f-j);
				\path (h f-j) edge[bend right,above] node{$j\xi$} (none40);
				\path (none31) edge[bend right,left] node{$\eta$} (h f-j);
				\path (h f-j) edge[bend right,right] node{$h\xi$} (none31);
				
				\node[state, draw=none, minimum size=1.6cm] (none42) [right of=h f-j] {$\ldots$};
				\path (h f-j) edge[bend right,below] node{$\eta$} (none42);
				\path (none42) edge[bend right,above] node{$(j+1)\xi$} (h f-j);
				
				\node[state, minimum size=1.6cm] (h f) [right of=none42] {$h,f$};
				\path (none42) edge[bend right,below] node{$\eta$} (h f);
				\path (h f) edge[bend right,above] node{$f\xi$} (none42);
				\path (none32) edge[bend right,left] node{$\eta$} (h f);
				\path (h f) edge[bend right,right] node{$h\xi$} (none32);
				
	\end{tikzpicture}%
    }
    \caption{A representation of the availability model for a BFT consensus protocol.}
    \label{fig:availability_model}
\end{figure}